\documentclass{emulateapj}

\usepackage{natbib}

\shorttitle{Galaxy Environments at $z \sim 1$}
\shortauthors{Cooper et al.}

\begin{document}


\title{The DEEP2 Galaxy Redshift Survey: The Relationship between
Galaxy Properties and Environment at \boldmath$\lowercase{z} \sim 1$}


\author{
Michael C.\ Cooper\altaffilmark{1},
Jeffrey A.\ Newman\altaffilmark{2,3},
Darren J.\ Croton\altaffilmark{1},
Benjamin J.\ Weiner\altaffilmark{4},
Christopher N.\ A.\ Willmer\altaffilmark{4},
Brian F.\ Gerke\altaffilmark{5},
Darren S.\ Madgwick\altaffilmark{2,3},
S.\ M.\ Faber\altaffilmark{4},
Marc Davis\altaffilmark{1,5},
Alison L.\ Coil\altaffilmark{6,3},
Douglas P.\ Finkbeiner\altaffilmark{7},
Puragra Guhathakurta\altaffilmark{4},
David C.\ Koo\altaffilmark{4}
}

\altaffiltext{1}{
Department of Astronomy, University of California at Berkeley, Mail Code
3411, Berkeley, CA 94720 USA; cooper@astro.berkeley.edu,
darren@astro.berkeley.edu, marc@astro.berkeley.edu}

\altaffiltext{2}{
Lawrence Berkeley National Laboratory, 1 Cyclotron Road Mail Stop 50-208,
Berkeley, CA 94720 USA; janewman@lbl.gov, dsmadgwick@lbl.gov}

\altaffiltext{3}{Hubble Fellow}

\altaffiltext{4}{UCO/Lick Observatory, UC Santa Cruz, Santa Cruz, CA
95064 USA; bjw@ucolick.org, cnaw@ucolick.org, faber@ucolick.org, 
raja@ucolick.org, koo@ucolick.org}

\altaffiltext{5}{
Department of Physics, University of California at Berkeley, Mail Code
7300, Berkeley, CA 94720 USA; bgerke@astro.berkeley.edu}

\altaffiltext{6}{ 
Steward Observatory, University of Arizona, 933 N.\ Cherry Avenue, 
Tucson, AZ 85721 USA; acoil@as.arizona.edu}

\altaffiltext{7}{
Princeton University Observatory, Princeton, NJ 08544 USA;
dfink@astro.princeton.edu}

\begin{abstract}

We study the mean environment of galaxies in the DEEP2 Galaxy Redshift
Survey as a function of rest-frame color, luminosity, and [OII]
3727\AA\ equivalent width. The local galaxy overdensity for $> \!
14,000$ galaxies at $0.75 < z < 1.35$ is estimated using the projected
$3^{\rm rd}$-nearest-neighbor surface density. Of the galaxy
properties studied, mean environment is found to depend most strongly
on galaxy color; all major features of the correlation between mean
overdensity and rest-frame color observed in the local universe were
already in place at $z \sim 1$. In contrast to local results, we find
a substantial slope in the mean dependence of environment on
luminosity for blue, star-forming galaxies at $z \sim 1$, with
brighter blue galaxies being found on average in regions of greater
overdensity. We discuss the roles of galaxy clusters and groups in
establishing the observed correlations between environment and galaxy
properties at high redshift, and we also explore the evidence for a
``downsizing of quenching'' from $z \sim 1$ to $z \sim 0$. Our results
add weight to existing evidence that the mechanism(s) that result in
star-formation quenching are efficient in group environments as well
as clusters. This work is the first of its kind at high redshift and
represents the first in a series of papers addressing the role of
environment in galaxy formation at $0 < z < 1$.

\end{abstract}


\section{Introduction}
For more than a few decades now, the observed properties of galaxies
have been known to depend upon the environment in which they are
located. For instance, elliptical and lenticular galaxies are
systematically over-represented in highly overdense environments such
as galaxy clusters \citep[e.g.][]{davis76, dressler80, postman84,
balogh98}. Similarly, measurements of the 2-point galaxy correlation
function for different luminosity and color subpopulations have
provided a complementary window to what galaxies populate different
environments \citep{zehavi02, norberg02, madgwick03,
coil04a}. Furthermore, recent studies utilizing the Sloan Digital Sky
Survey \citep[][SDSS]{york00} and the 2-degree Field Galaxy Redshift
Survey \citep[][2dFGRS]{colless01} have found that a wide array of
galaxy properties (e.g.\ color, luminosity, surface-brightness,
star-formation rate, and AGN activity) correlate with the local
density of the galaxy environment and that the observed correlations
extend from the centers of clusters out into the field galaxy
population \citep{lewis02, gomez03, balogh04, hogg04, christlein05}.

Current hierarchical galaxy formation models predict that galaxies
form in less dense environments and are then accreted into larger
groups and clusters \citep{kauffmann93, somerville99, cole00,
delucia05}. Within such models, there are a plethora of physical
processes that could explain the observed trends with environment. For
example, galaxy mergers, which preferentially occur in galaxy groups
rather than clusters \citep[e.g.][]{cavaliere92}, can destroy galactic
disks and thereby convert spiral galaxies into ellipticals and
lenticulars (Toomre \& Toomre 1972); likewise, ram-pressure stripping
within galaxy clusters can effectively suffocate star formation in
member galaxies (Gunn \& Gott 1972), and within overdense regions
galaxies can be swept of their gas via feedback from AGN, thus
quenching the star-formation process \citep{springel05}.

In recent years, many studies have focused on measuring the
correlations between galaxy properties and environment at $z \sim
0$. For example, using large data sets from local surveys such as the
SDSS, \citet{blanton04}, \citet{kauffmann04}, \citet{croton05a}
measured the relationship between star-formation history and local
galaxy density for environments ranging from voids to clusters. At
high redshift, however, such analysis over a similar environment range
and with a comparable sample size has yet to be completed due in large
part to the lack of a suitable data set. While studies of clusters
have pushed to high redshift, the range of environments probed in such
studies has been limited, with galaxies grouped into coarse
classifications (such as field, group, and cluster populations) and
with statistically small spectroscopic sample sizes
\citep[e.g.,][]{couch98, treu03, smith05, tanaka05, poggianti05}. As
discussed by \citet{cooper05}, to measure galaxy environments
accurately at $z \sim 1$, a survey must obtain high-precision (i.e.\
spectroscopic) redshifts over a large and contiguous field with a
relatively high sampling rate. The DEEP2 Galaxy Redshift Survey
\citep{davis03, faber06} is the first survey at high redshift to
provide such a sample, opening the door to studying galaxy properties
and environments at $z \sim 1$ over a continuous range of environments
from voids to large groups.

Studying the relationship between environment and galaxy properties at
high redshift should afford the perspective needed to understand the
nature of the observed relations found locally. In particular,
studying galaxy environments at higher $z$ will help determine whether
the correlations observed in the nearby universe are a result of
evolutionary processes with long time-scales or whether the observed
relations were in place very early in the lifetime of the
universe. Constraints derived from $z \sim 1$ studies should also
yield information regarding the physical processes (e.g.\ ram-pressure
stripping, harassment, and mergers) and possible environment regimes
(e.g.\ groups versus clusters) that are significant in creating the
trends seen locally.

In this paper, we study the environment of high-redshift $(z \sim 1)$
galaxies in the DEEP2 Galaxy Redshift Survey with a goal of
determining the correlation between local density and galaxy
properties when the universe was half its present age. In particular,
we examine the relations between environment and galaxy color,
luminosity, and [OII] equivalent width. In \S2, we describe the DEEP2
survey and the galaxy sample under study. The measurements of galaxy
properties including color, luminosity, [OII] equivalent width, and
local density are dicussed in \S3 and \S4. In \S5, we present our
results and a qualitative comparison to observations at $z \sim
0$. Lastly, in \S6 we summarize and discuss our findings with some
additional discussion directed towards future analysis utilizing the
DEEP2 survey. Throughout this paper, we employ a $\Lambda$CDM
cosmology with $w = -1$, $\Omega_M = 0.3$, $\Omega_{\Lambda} = 0.7$,
and $h = 1$.

\section{Data Sample}

The DEEP2 Galaxy Redshift Survey \citep{davis03,faber06} is an ongoing
project designed to study the galaxy population and large-scale
structure at $z \sim 1$. The survey utilizes the DEIMOS spectrograph
\citep{faber03} on the 10-meter Keck II telescope to target $\gtrsim
\! 40,000$ galaxies covering $\sim 3$ square degrees of sky over four
widely separated fields. In each field, targeted galaxies are selected
to $R_{\rm AB} \le 24.1$ from deep $B,R,I$ imaging taken with the
CFH12k camera on the 3.6-meter Canada-France-Hawaii Telescope
\citep{coil04b, davis04}. Using a simple color-cut in three of the
survey fields, the high-redshift $(z > 0.7)$ galaxies are selected for
observation with only $\lesssim \! 3\%$ of galaxies at $z > 0.75$
rejected, based on tests in the fourth survey field, the Extended
Groth Strip \citep{davis04, faber06}.

This work utilizes data from the first six DEEP2 observing seasons
(collected from August 2002 -- July 2005) spread over three of the
survey's four fields. The DEIMOS data were processed using a
sophisticated IDL pipeline, developed at UC-Berkeley \citep{cooper06b}
and adapted from spectoscopic reduction programs created for the SDSS
\citep{burles06}. The observed spectra were taken at moderately high
spectral resolution $(R \sim 5000)$ and typically span an observed
wavelength range of $6300{\rm \AA}\ < \lambda < 9100{\rm
\AA}$. Working at such high resolution, we are able to unambigously
detect and resolve the [OII] 3727\AA\ doublet for galaxies in the
redshift range $0.7 < z < 1.4$. Absorption-based redshifts are also
measured, primarily relying upon the Ca H and Ca K features which are
detectable out to redshifts of $z \sim 1.3$.
Repeated observations of a subset of galaxies yield an effective
velocity uncertainty of $\sim \! 30\ {\rm km}/{\rm s}$ due to
differences in the position or alignment of a galaxy within a slit or
internal kinematics within a galaxy \citep{faber06}.

\begin{figure}[h!]
\begin{center}
\plotone{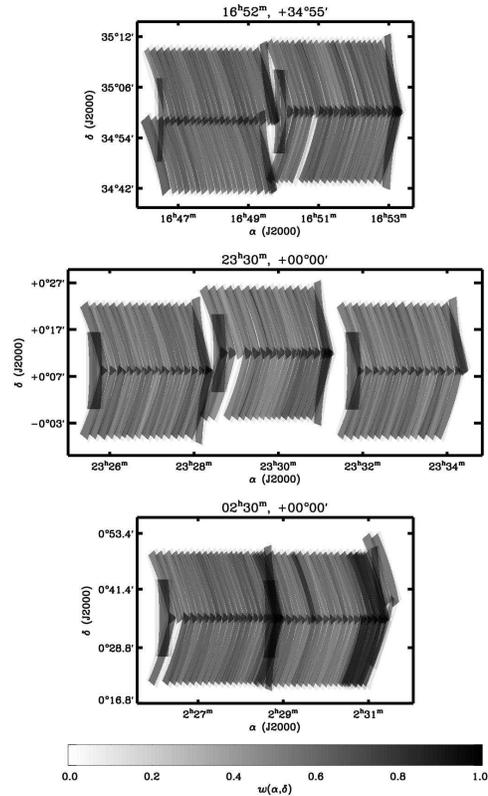}
\caption{We present the 2-d redshift completeness map, $w(\alpha,
\delta)$, for spectroscopic coverage in each survey field. We include
all 280 slitmasks with redshift completeness for red $({\rm observed}\
R-I > 0.5)$ galaxies $\ge 65\%$ in our analysis. The greyscale in each
image ranges from 0 (white) to 0.95 (black) and corresponds to the
probability that a galaxy meeting the survey selection criteria at
that position on the sky was targeted for spectroscopy and a redshift
was successfully measured. The color bar at the bottom shows the value
of $w(\alpha, \delta)$ as function of the greyscale.}
\label{wfn}
\end{center}
\end{figure}

\begin{figure}[h!]
\begin{center}
\plotone{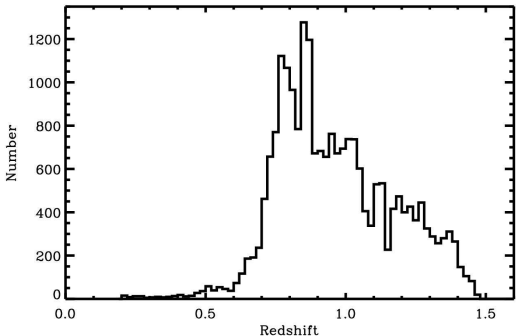}
\caption{The observed redshift distribution for all 23,004 sources in
the three surveyed regions. The color cut rejects $\lesssim 3\%$ of
galaxies with $z > 0.75$, based on tests in the Extended Groth Strip
\citep{faber06}. The redshift histogram is plotted using a bin size of
$\Delta z = 0.02$.}
\label{nofz}
\end{center}
\end{figure}

We present results from the most complete portions of three of the
DEEP2 fields. The data sample spans 280 DEIMOS slitmasks, covering a
total of $\gtrsim \! 2\ {\rm degrees}^2$ of sky with an average
redshift completeness of $\gtrsim 70\%$. We use data only from
slitmasks which have a redshift success rate of 65\% or higher for red
$(R-I > 0.5)$ galaxies in order to avoid systematic effects which may
bias our results. The success rate for red galaxies is well correlated
with the signal-to-noise of the 1-d spectra on a slitmask, with a
roughly Gaussian distribution of redshift completeness (that is,
success rate) above 65\%, with a mean of $\sim 80\%$. Figure \ref{wfn}
shows the 2-dimensional redshift completeness map for each of the
surveyed fields.
The completeness map gives the probability at that position on the sky
that a galaxy meeting the survey selection criteria was targeted for
spectroscopy and a redshift was successfully measured. The total data
sample (Sample A in Table \ref{sample_tab}) in the three fields
includes 22,416 sources with accurate redshift determinations
\citep[quality $Q=3$ or $Q=4$ as defined by][]{faber06} within the
redshift range $0.2 < z < 2$; we show the redshift distribution of
this sample in Figure \ref{nofz}.

\section{Measurements of Galaxy Properties}

In this section, we describe the set of galaxy properties for which we
study the distributions and correlations with local environment. The
properties to be discussed include galaxy color, luminosity, and [OII]
equivalent width. In the following subsections, we detail the
measurement of each property within the DEEP2 sample.

\subsection{Galaxy Colors and Luminosities}

The rest-frame color, $(U-B)_0$, and absolute $B$-band magnitude,
$M_B$, are calculated by combining the CFHT $B,R,I$ photometry with
spectral energy distributions (SEDs) of galaxies ranging from 1180\AA\
to 10000\AA\ as compiled by \cite{kinney96}. By projecting the
\cite{kinney96} SEDs onto the response functions for the CFHT
filters\footnote{The filter response curves are available for download
at http://deep.berkeley.edu/DR1/ or
http://www.cfht.hawaii.edu/Instruments/Filters/filterdb.html.} and for
the corresponding standard Johnson filters \citep{bessell90}, we
compute the rest-frame $U-B$ and $B-R_{\rm cfht}$
colors,\footnote[1]{Note that in this section the CFHT filters are
differentiated from the Johnson filter set according to the subscript
notation.} the ${\rm K}_{R_{\rm cfht}}$ transformations, and the
expected observed colors for each SED at the redshift of the galaxy
for which we are correcting. A low-order polynomial is then fit
between the synthetic $B_{\rm cfht} - R_{\rm cfht}$, $R_{\rm cfht} -
I_{\rm cfht}$ colors and the $U-B$ color and $B-R_{\rm cfht} - {\rm
K}_{R_{\rm cfht}}$ transformation. By using the coefficients of the
fit, the final rest-frame color and absolute magnitude for each galaxy
are calculated in the Johnson filter set. In computing the
K-corrections, we follow the notation of \citet{hogg02}. For all
details regarding the DEEP2 photometric catalog and measured source
fluxes, refer to \citet{coil04b}. For specifics relating to the
computation of rest-frame colors and luminosities, refer to
\citet{willmer05}. Unless explicitly stated otherwise, all magnitudes
discussed in this paper are given in AB magnitudes \citep{oke83}. For
zero-point conversions between AB and Vega magnitudes, refer to Table
1 of \citet{willmer05}.

\begin{figure}[h!]
\begin{center}
\plotone{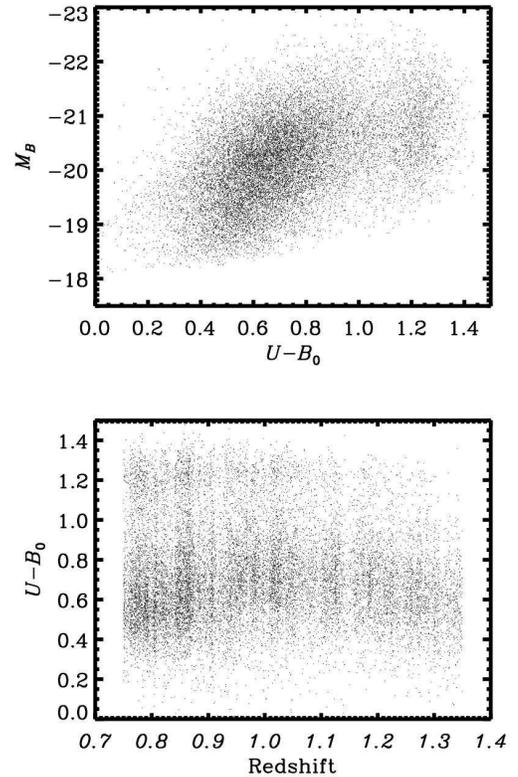}
\caption{({\emph Top}) Plotted is the color-magnitude relation for the
  18,977 sources in the sample within the redshift range $0.75 < z <
  1.35$ (Sample B in Table \ref{sample_tab}). ({\emph Bottom}) We plot the
  rest-frame $(U-B)_0$ color distribution with redshift for the same
  sample. At high redshift $(z > 1.1)$, the red galaxy fraction drops
  significantly due primarily to the $R_{\rm AB}$ magnitude limit of
  the survey \citep{willmer05}.}
\label{cmd}
\end{center}
\end{figure}

Figure \ref{cmd} shows the color-magnitude relation for the 18,977
galaxies at $0.75 < z < 1.35$ in the data sample. A distinct
bimodality exists in the color distribution with a division at
$(U-B)_0 \sim 1.0$, separating members of the ``blue cloud'' from
red-sequence galaxies
\citep[e.g.][]{strateva01,baldry04,bell04,weiner05}. Within the DEEP2
sample, the division between the red and blue populations shows little
sign of evolution \citep{willmer05}; as illustrated in Figure
\ref{cmd}, over the entire redshift range probed in the data set, a
strong division is observed at $(U-B)_0 \sim 1.0$. However, at
redshifts beyond $z \sim 1.1$, the red galaxy fraction decreases
precipitously in the DEEP2 sample, due to the survey's $R$-band
magnitude limit. For a more thorough discussion of rest-frame colors
and luminosities in the DEEP2 survey, we direct the reader to
\citet{willmer05}.

\subsection{[OII] Equivalent Widths}
For each object in the sample, we measure the equivalent width of the
[OII] 3727\AA\ doublet if it is within the wavelength coverage of the
observed spectrum. We utilize the boxcar-extracted 1-D spectrum and
errors produced by the DEEP2 pipeline \citep{cooper06b, faber06} and
fit a double Gaussian profile to the wavelength, flux, and error of
pixels in a 40\AA\ window centered on the predicted location of the
[OII] emission. The fit uses a Levenberg-Marquardt nonlinear least
squares minimization routine \citep{press92} with the following free
parameters: continuum, intensity, central wavelength, dispersion, and
intensity ratio of the two lines in the [OII] doublet. This routine
produces best-fit values and error estimates. In fitting the Gaussian
profiles, the wavelength ratio of the two lines in the doublet is
fixed, but the intensity ratio is allowed to vary. The fitted
continuum is noisy when the data have low S/N, so we make a robust
estimate of the continuum by taking the biweight \citep{beers90} of
all data in two windows on either side of the line, each 80\AA\ long
and separated from the line by a 20\AA\ buffer. The rest-frame [OII]
equivalent width and error are derived from the total intensity of the
doublet and the robust continuum. For all analysis utilizing [OII]
equivalent widths, we limit the sample studied to those galaxies for
which the uncertainty in $W_{\rm [OII]}$ is small $(\sigma_{W_{\rm
[OII]}} < 10\ {\rm \AA})$, a selection that retains $\sim 45\%$ of
galaxies.

\section{Local Galaxy Densities}

For each galaxy in the data set, we estimate the local galaxy density
using the projected $3^{\rm rd}$-nearest-neighbor surface density,
$\Sigma_{3}$. In computing $\Sigma_3$, the full DEEP2 galaxy sample is
employed and we utilize a velocity interval of $\pm 1000\ {\rm
km}/{\rm s}$ along the line-of-sight to exclude foreground and
background galaxies. The measured surface density depends on the
projected distance to the $3^{\rm rd}$-nearest neighbor, $D_{p,3}$, as
$\Sigma_3 = 3 / \left( \pi D_{p,3}^2 \right)$, such that $\Sigma_3 =
0.1\ {\rm galaxies}/{(h^{-1} {\rm comoving Mpc})^2}$ corresponds to a
projected $3^{\rm rd}$-nearest-neighbor distance of $\sim 3\ h^{-1}$
comoving Mpc and surface density of unity corresponds to a length
scale of $D_{p,3} \sim 1\ h^{-1}$ comoving Mpc. From tests using the
mock galaxy catalogs of \citet{yan04}, \citet{cooper05} find the
projected $3^{\rm rd}$-nearest-neighbor distance to be a robust and
accurate environment measure that minimizes the role of redshift-space
distortions and edge effects. Further details regarding the
computation and sensitivity of this density measure as well as
comparisons to other common environment estimators are presented in
\citet{cooper05}.

\subsection{Correcting for Variations in Redshift Sampling and
Completeness} 

To correct each density estimate for the variable redshift
completeness of the DEEP2 survey, we scale the density measured about
each galaxy according to the local value of the 2-dimensional survey
completeness map, $w(\alpha, \delta)$, which accounts for variations
in targeting rate and redshift completeness from position to position
within the survey \citep{coil04a}. Each density value is also
corrected for the redshift dependence of the sampling rate of the
survey using the empirical method presented in \citet{cooper05}: each
density value is divided by the median $\Sigma_3$ of galaxies at that
redshift over the whole sample, where the median is computed in bins
of $\Delta z = 0.04$. Using mock galaxy catalogs, \citet{cooper05}
conclude that such an empirical correction is effective at reducing
the influence of variations in redshift sampling in the survey without
overcorrecting variations in enviroment due to cosmic
variance. Correcting the measured densities in this manner converts
the $\Sigma_3$ values into measures of overdensity relative to the
median density $({\rm denoted}\ {\rm by}\ 1 + \delta_3)$, and is similar
to the methods employed by \citet{hogg03} and \citet{blanton04}. To
ensure that the applied corrections are not exceedingly large, we
restrict the analysis presented in this paper to galaxies in the
redshift interval $0.75 < z < 1.35$, where the DEEP2 selection
function is relatively shallow.

\subsection{Edge Effects}
When calculating the local density of galaxies, the confined sky
coverage of a survey introduces geometric distortions -- or edge
effects -- which bias environment measures near edges or holes in the
survey field, generally causing densities to be under-estimated
\citep{cooper05}. We define the DEEP2 survey area and corresponding
edges according to the 2-dimensional survey completeness map and the
photometric bad-pixel mask used in slitmask design -- defining all
regions of sky with $w(\alpha, \delta) < 0.35$ averaged over scales of
$\gtrsim 30''$ to be unobserved and rejecting all significant regions
of sky ($\gtrsim 30''$ in scale) which are incomplete in the
photometric catalog. Areas of incompleteness on scales $\lesssim 30''$
are comparable to the typical angular separation of galaxies targeted
by DEEP2 \citep{gerke05, cooper05}, and thus cause a negligible
perturbation to the measured densities.

\begin{figure}[h!]
\begin{center}
\plotone{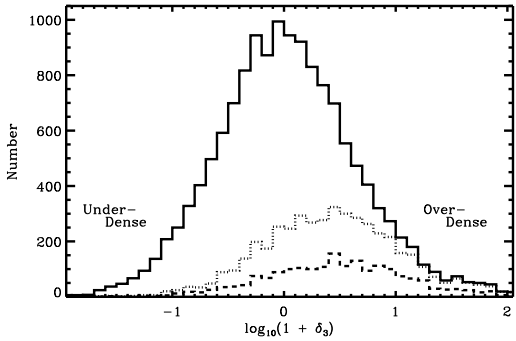}
\caption{In logarithmic space, we plot the distribution of local
overdensities, $(1 + \delta_3)$, for the 14,214 galaxies with $0.75 <
z < 1.35$ and more than $1\ h^{-1}$ comoving Mpc from a survey edge,
comprising the Sample C (see Table \ref{sample_tab}) for this paper
(\emph{solid line}). We also show the distribution of overdensities
for galaxies in groups and clusters as identified by
\citet{gerke05}. The \emph{dotted line} gives the distribution of
$\log_{10}{(1 + \delta_3)}$ for all galaxies in groups or clusters and
the \emph{dashed line} gives the distribution for galaxies in groups
or clusters with a velocity dispersion $\sigma_{\rm group} > 200\ {\rm
km}/{\rm s}$ as measured by \citet{gerke05}. The overdensity,
$(1 + \delta_3)$, is a dimensionless quantity, computed as described in
\S4.1.}
\label{envdist}
\end{center} 
\end{figure}

To minimize the impact of edges on the data sample, we exclude any
galaxy within $1\ h^{-1}$ comoving Mpc of an edge or gap. Such a cut
greatly reduces the portion of the data set contaminated by edge
effects \citep{cooper05} while retaining $\sim \! 75\%$ of the full
galaxy sample. Combined with the restriction to the redshift regime
$0.75 < z < 1.35$, this gives a final sample comprising 14,214
galaxies. For easy reference, the details regarding the galaxy samples
utilized throughout this paper are listed in Table
\ref{sample_tab}. The distribution of overdensities, $(1 + \delta_3)$,
for this sample (Sample C in Table \ref{sample_tab}) is shown in
Figure \ref{envdist}.

\subsection{Target-Selection Effects}

Due to the need to design DEIMOS slitmasks such that spectra do not
overlap and due to the adopted target-selection and slitmask-tiling
schemes of the survey, the DEEP2 survey slightly under-samples
projected overdensities of galaxies on the sky \citep{gerke05,
cooper05}. Such a bias in the survey is a critical concern for the
measurement of accurate galaxy densities in highly-clustered
environments. However, using the mock catalogs of \citet{yan04} to
test the survey target-selection and slitmask-making procedures,
\citet{cooper05} find that while the DEEP2 survey under-samples dense
\emph{regions of sky} (projected densities on the sky), the survey
does not under-sample dense \emph{environments} (that is, local
densities in three-dimensional space) to any significant degree.

\subsection{Measurement Errors}

We have determined the uncertainty in our environment measures using
the mock galaxy catalogs of \citet{yan04}. In order to simulate the
DEEP2 sample, the volume-limited mock catalogs are passed through the
DEEP2 target-selection and slitmask-making procedures as described by
\citet{davis04} and \citet{faber06}, placing $\sim 60\%$ of targets on
a slitmask. Using the same techniques as applied to the data, we
measure the environment for each galaxy in the simulated DEEP2
sample. For each mock galaxy, this ``observed'' environment is
compared to the ``true'' environment as measured in the full
volume-limited catalog using the real-space positions for all of the
galaxies. Measuring the scatter in ``true'' minus ``observed''
environment for the simulated DEEP2 galaxies shows a 1-$\sigma$
scatter of roughly 0.5 in units of $\log{(1 + \delta_3)}$. The uncertainty
in the overdensity values shows little dependence on environment, with
only a slight increase in the errors at higher densities. For more
details on the mock catalogs or tests of environment measures using
them, refer to \citet{yan04} and \citet{cooper05}, respectively.

\section{Results}

The local density of galaxies is thought to influence galaxy
characteristics such as morphology and star-formation rate, and thus
such galaxy properties are often studied as a function of
enviroment. Analyses performed in this manner are particularly helpful
at recognizing scales or densities at which galaxy properties
transition and are therefore useful in understanding the physical
processes at play. Measurements of galaxy environment, however, are
significantly more uncertain than measures of other properties such as
color or luminosity. Therefore, by binning galaxies according to local
density, there is a significant correlation between neighboring
density bins which smears out any trends with other galaxy
properties. Consequently, in this paper, we instead study the
dependence of mean environment on galaxy properties. The mean
relations are weighted according to the inverse selection volumes, $1
/ V_{\rm max}$; computing the mean in this manner minimizes the
effects of Malmquist bias \citep{malmquist36} in the sample. In
computing the $1 / V_{\rm max}$ values, we restrict the surveyed
volume to $0.75 < z < 1.35$ and incorporate rest-frame color according
to the K-corrections discussed in \S3.1.

\subsection{Dependence of Mean Environment on Galaxy Color}

\begin{figure*}[h]
\begin{center}
\plotone{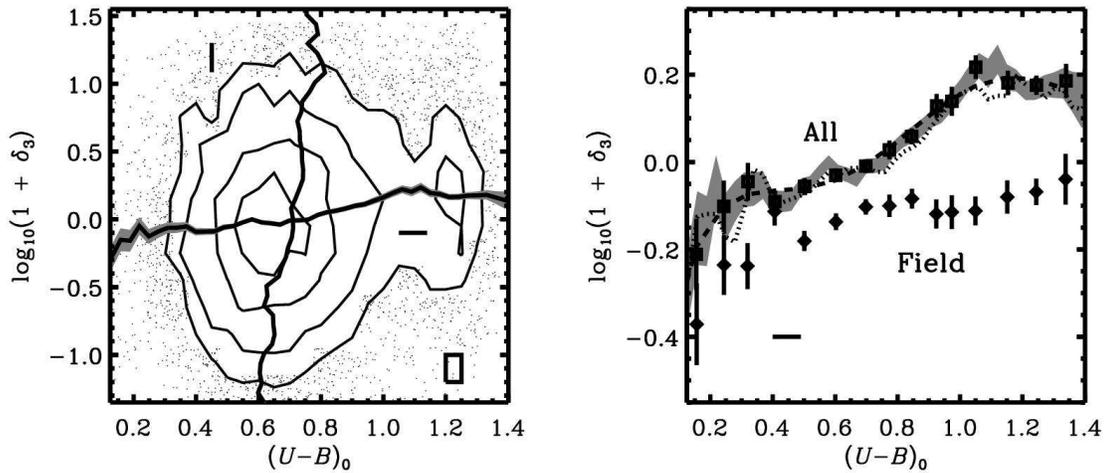}
\caption{(\emph{Left}) We plot the logarithm of the local galaxy
overdensity, $(1 + \delta_3)$, versus the rest-frame color, $(U-B)_0$,
for all 14,214 galaxies in the $0.75 < z < 1.35$ sample (Sample
C). The contours show the number density of sources as a function of
$\log{(1 + \delta_3)}$ and $(U-B)_0$ and correspond to levels of 25,
50, 100, and 150 galaxies. The number density is computed in a sliding
box of height $\Delta \log{(1 + \delta_3)} = 0.2$ and width $\Delta
(U-B)_0 = 0.05$ as illustrated in the lower right corner of the
plot. The solid black horizontal line shows the weighted mean
dependence of environment on color, while the vertical counterpart
gives the weighted mean dependence of $(U-B)_0$ on environment. The
respective means were computed using sliding boxes with widths given
by the black dashes in the plot and were weighted according to the
inverse selection volumes, $1/V_{\rm max}$. The accompanying grey
regions correspond to the sliding 1-$\sigma$ uncertainties in the
weighted means. (\emph{Right}) We again show the 1-$\sigma$ region for
the weighted mean dependence of environment on galaxy color computed
in a sliding box of width given by the black dash (grey region). The
square points and error bars give the weighted mean environments and
1-$\sigma$ uncertainties in the means computed in distinct bins of
color, thereby avoiding the covariance associated with the sliding
mean.
The diamond points and error bars show the weighted mean environments
and 1-$\sigma$ uncertainties in the means computed in indentical color
bins for the field galaxy population solely. The field population is
selected according to the galaxy group finder of \citet{gerke05},
excluding all galaxies identified as group members. The dotted line
shows the median $\log_{10}{(1 + \delta_3)}$ versus color trend utilizing
the same sliding box. Lastly, a $6^{\rm th}$-order polynomial fit to
the mean color dependence of galaxy environment is shown by the dashed
line and used below (see Table \ref{fits_tab} for coefficients of the
fit). At $z \sim 1$, we find the relationship between rest-frame color
and environment to echo the trend measured locally with red galaxies
favoring regions of high overdensity and very blue objects
preferrentially found in underdense environments.}
\label{environ_vs_color}
\end{center}
\end{figure*}

As shown in Figure \ref{environ_vs_color}, at $z \sim 1$, blue
galaxies on average occupy less-dense environments. We observe a
distinct transition in mean overdensity which corresponds well with
the observed color bimodality. The mean environment for red galaxies
at $z \sim 1$ is more than 1.5 times more dense than for their blue
counterparts. At very blue colors, $(U-B)_0 < 0.3$, we also find
significant evidence for a fall-off in the mean overdensity.

The observed relationship between galaxy overdensity and rest-frame
color at $z \sim 1$ mirrors that seen in the local universe. The
dependence of mean environment on color at $z \sim 0.1$ exhibits a
strong transition in mean density occurring at the location of the
color bimodality and a further decrease in mean density for the very
bluest galaxies \citep{hogg03, blanton03b, balogh04b}. We find that
this trend is already in place as early as a redshift of unity.

As illustrated in Figure \ref{cmd}, the DEEP2 sample is heavily
weighted towards blue galaxies at higher redshift; in addition, the
sampling rate for DEEP2 decreases significantly beyond $z > 1.1$ (see
Fig.\ \ref{nofz}). The observed correlation between mean overdensity
and color, however, is not a product of selection effects. By limiting
the sample to $0.75 < z < 1.05$ and $M_B < -20$ where the red galaxy
sample is complete and any evolution in the measured galaxy density is
negligible, the observed color dependence of mean environment persists
with the contrast in typical overdensity for red and blue galaxies
unchanged relative to that observed for the full $0.75 < z < 1.35$
sample.

Using the group finder of \citet{gerke05}, we are able to identify the
set of galaxies within the data set not associated with groups or
clusters, that is, the field population. In Figure
\ref{environ_vs_color}, we show the mean dependence of environment on
rest-frame $(U-B)_0$ color for this sub-population (diamond
symbols). We find that among the field sample a weak trend with color
exists such that red field members are found to be only slightly
overdense relative to their blue conterparts. At very blue colors,
however, the trend with mean overdensity is clearly evident within the
field galaxy population.

\subsection{Dependence of Mean Environment on [OII] Equivalent Width}

\begin{figure*}[h]
\begin{center}
\plotone{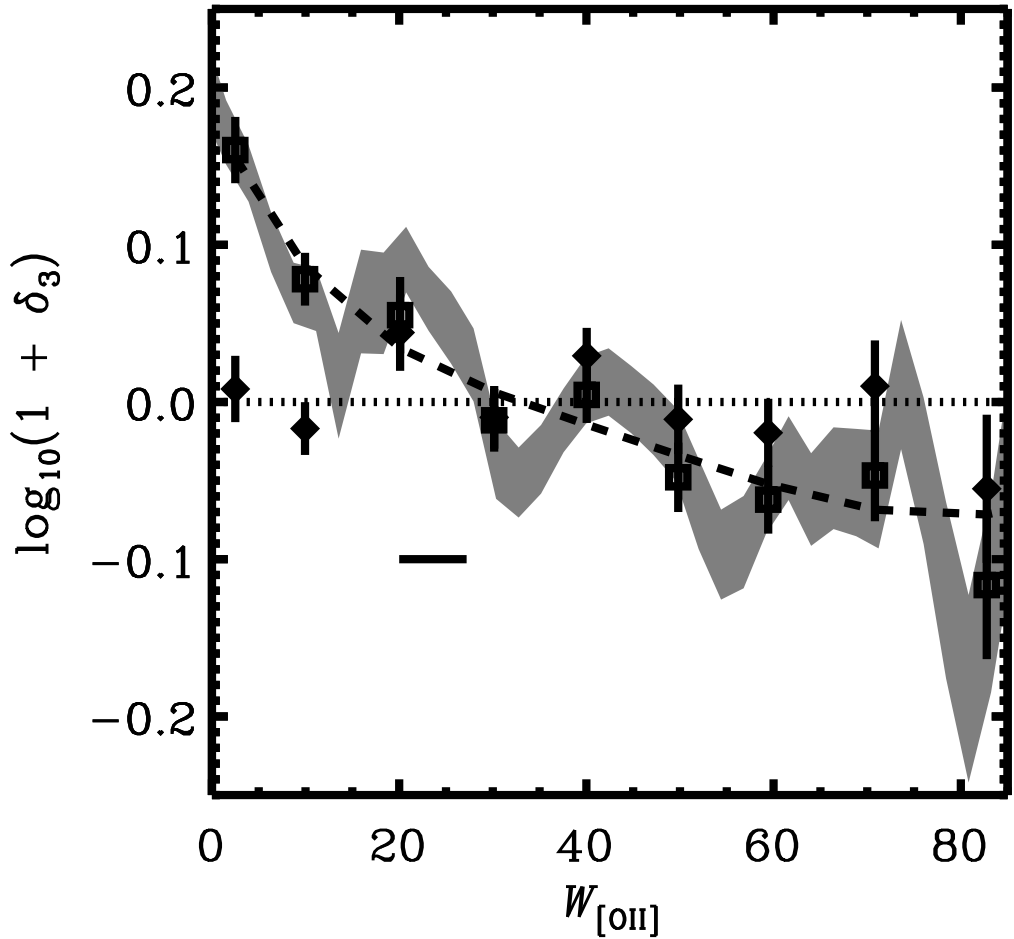}
\caption{We plot the mean relationship between the logarithm of the
local galaxy overdensity, $(1 + \delta_3)$, and the measured [OII]
equivalent width, $W_{\rm [OII]}$, for all 9,567 galaxies in the
sample with $\sigma_{W_{\rm [OII]}} < 10 {\rm \AA}$ (Sample D). We
compute the weighted mean dependence of environment on [OII]
equivalent width, using a sliding box with width given by the black
dash in the plot and weighted according to the inverse selection
volumes, $1/V_{\rm max}$. The dashed line follows a $5^{\rm th}$-order
polynomial fit to this mean dependence of environment on [OII]
equivalent width. The plotted grey region corresponds to the sliding
1-$\sigma$ uncertainty in the weighted mean. The square points and
error bars give the weighted mean environments and 1-$\sigma$
uncertainties in the means computed in distinct bins of $W_{\rm
[OII]}$, thereby avoiding the covariance associated with the sliding
mean. We fit a $6^{\rm th}$-order polynomial to the mean density
relation with color (see Fig.\ \ref{environ_vs_color}) and subtract
off this color dependence for each galaxy according to its rest-frame
color. The diamond points and error bars show the weighted mean
environments and 1-$\sigma$ uncertainties in the means computed in
distinct bins of $W_{\rm [OII]}$ after the mean color dependence has
been removed.
We find a strong relationship between [OII] equivalent width and mean
environment at $z \sim 1$, but we detect no additional dependence on
$W_{\rm [OII]}$ separate from the dependence on rest-frame color.}
\label{environ_vs_ew}
\end{center}
\end{figure*}

As shown in Figure \ref{environ_vs_ew}, the dependence of mean galaxy
density on [OII] equivalent width, $W_{\rm [OII]}$, is a strong
monotonic trend at $z \sim 1$. On average, galaxies with smaller
equivalent widths occupy regions of higher density. Again, limiting
the sample to galaxies in the redshift range $0.75 < z < 1.05$ and
brighter than $M_B = -20$, we find the measured trend with $W_{\rm
[OII]}$ to be essentiallly unchanged; we do not appear to be biased by
any redshift dependence or selection effect in the sample.

\begin{figure}[h]
\begin{center}
\plotone{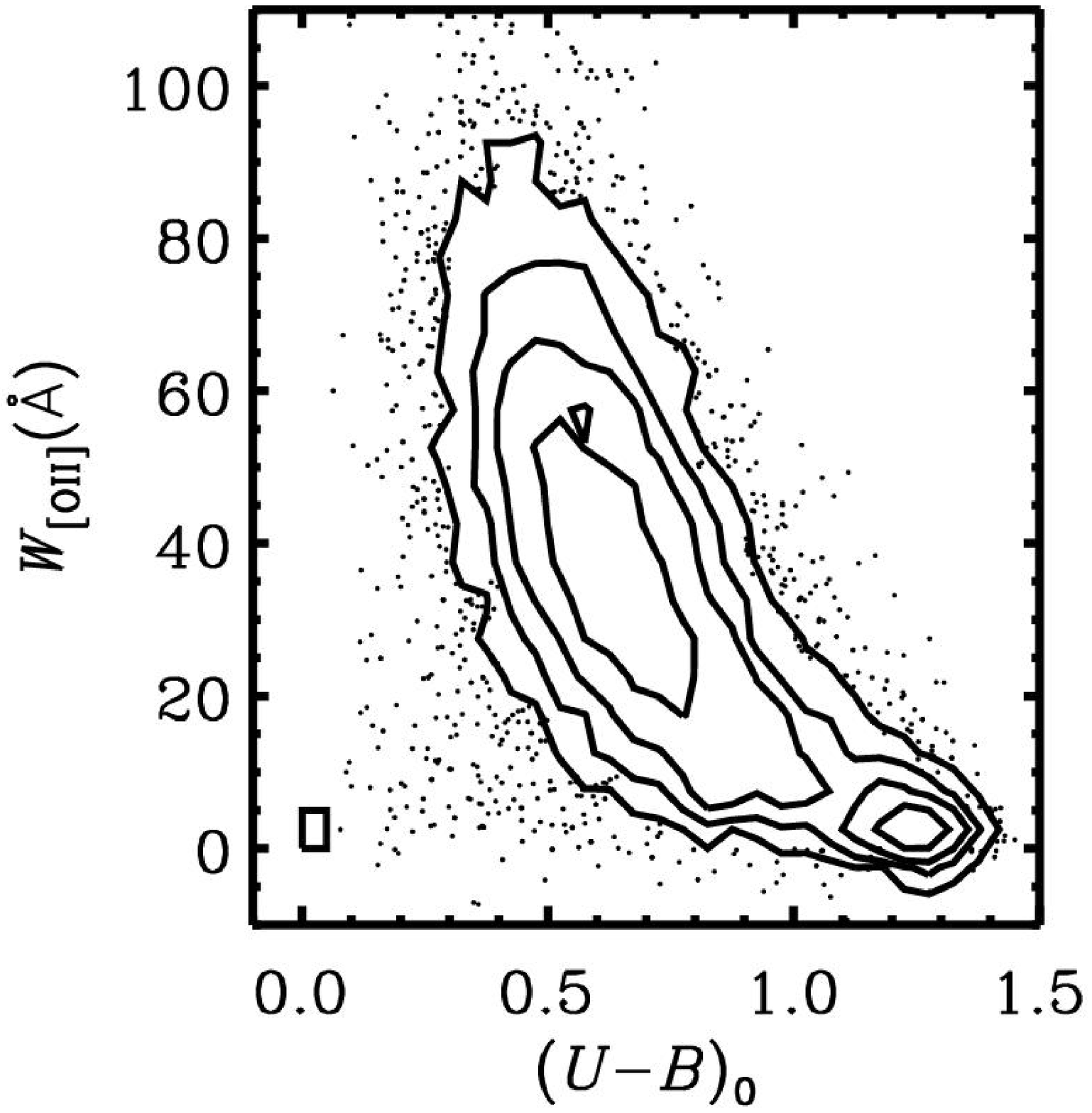}
\caption{The correlation between [OII] equivalent width, $W_{\rm
[OII]}$, and rest-frame galaxy color, $(U-B)_0$, over the randshift
range $0.75 < z < 1.35$. Here, we limit the sample plotted to those
galaxies comprising Sample D of Table \ref{sample_tab}. The contours
show the number density of sources as a function of $W_{\rm [OII]}$
and $(U-B)_0$ and correspond to levels of 10, 25, 50, and 100
galaxies. The number density is computed in a sliding box of height
$\Delta W_{\rm [OII]} = 5 {\rm \AA}$ and width $\Delta (U-B)_0 = 0.05$
as illustrated in the lower left corner of the plot.}
\label{ew_vs_color}
\end{center}
\end{figure}

\begin{figure*}[h]
\begin{center}
\plotone{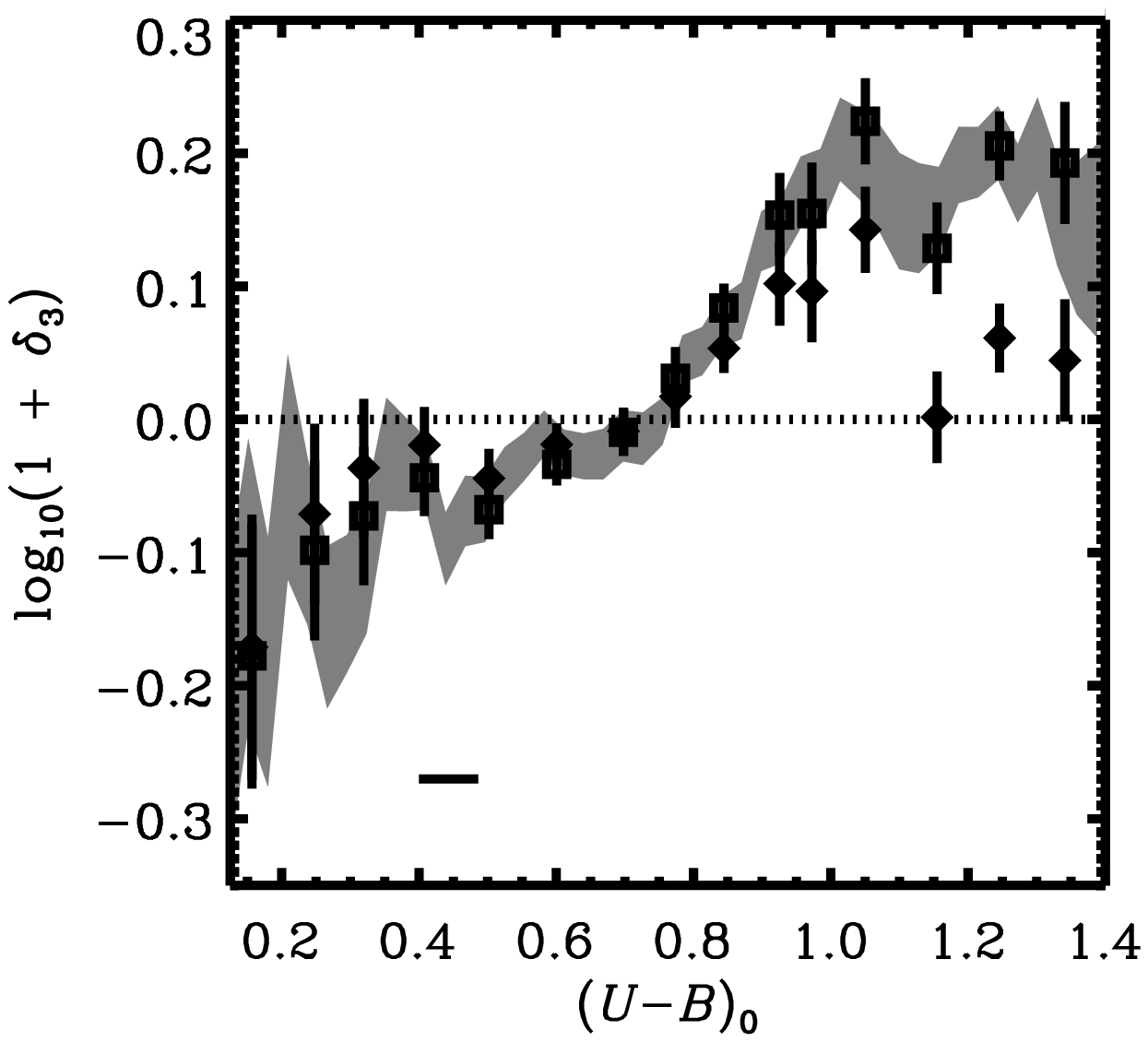}
\caption{We plot the relationship between the logarithm of the local
galaxy overdensity, $(1 + \delta_3)$, and the rest-frame color,
$(U-B)_0$ for all 9,567 galaxies in the sample with $\sigma_{W_{\rm
[OII]}} < 10 {\rm \AA}$ (Sample D). We compute the weighted mean
dependence of environment on color, using a sliding box with width
given by the black dash in the plot and weighted according to the
inverse selection volumes, $1/V_{\rm max}$. The plotted grey region
corresponds to the sliding 1-$\sigma$ uncertainty in the weighted
mean. The square points and error bars give the weighted mean
environments and 1-$\sigma$ uncertainties in the means computed in
distinct bins of $(U-B)_0$, thereby avoiding the covariance associated
with the sliding mean. We fit a $5^{\rm th}$-order polynomial to the
mean density relation with color (see Fig.\ \ref{environ_vs_ew} in the
paper) and subtract off this color dependence for each galaxy
according to its $W_{\rm [OII]}$. The diamond points and error bars
show the weighted mean environments and 1-$\sigma$ uncertainties in
the means computed in distinct bins of color after the mean color
dependence has been removed.
We find a strong relationship between color and mean
environment at $z \sim 1$, and we detect an additional dependence on
$(U-B)_0$ separate from the dependence on [OII] equivalent width.}
\label{environ_vs_color2}
\end{center}
\end{figure*}

The strength of the observed correspondence between environment and
$W_{\rm [OII]}$ is not surprising given the close correlation between
galaxy color and [OII] equivalent width, as shown in Figure
\ref{ew_vs_color} and found in previous studies of galaxy properties
at $z \sim 1$ \citep[e.g.][]{weiner05}. Red galaxies in the DEEP2
sample are strongly clustered in the $(U-B)_0, W_{\rm [OII]}$ plane
with $W_{\rm [OII]} < 40{\rm \AA}$, while for the blue galaxy
population [OII] equivalent width tends to increase rapidly with
decreasing color. Fitting a $5^{\rm th}$-order polynomial to the mean
overdensity relation with color (see Figure \ref{environ_vs_color})
and subtracting off this dependence for each galaxy according to its
rest-frame color, we find no evidence for any residual environment
dependence on [OII] equivalent width separate from the color
dependence; the residual points in Figure \ref{environ_vs_ew} are
consistent with a line of zero slope and intercept. Given the
uncertainties in the measured quantities, simulations indicate that we
would be able to detect a residual slope as small as $\sim \! 0.3
\log{(1 + \delta_3)} / {\rm \AA}$ equivalent width at a 3-$\sigma$
level. Conversely, if we apply this same test in reverse and check for
residuals in mean environment with $(U-B)_0$ color when the trend
between mean environment and $W_{\rm [OII]}$ has been removed, we find
that a strong residual trend between $<\log_{10}{(1 + \delta_3)}>$ and
$(U-B)_0$ does exist (see Fig.\ \ref{environ_vs_color2}). That is,
fitting to the dependence of mean overdensity on $W_{\rm [OII]}$ and
subtracting off this dependence for each galaxy according to its [OII]
equivalent width, we find a residual trend between mean environment
and color.

This result suggests that [OII] equivalent width does not have a
relationship with environment separate from that observed with
rest-frame $(U-B)_0$ color. While both galaxy properties are indirect
tracers of star-formation and local studies find a galaxy's
star-formation history to be strongly correlated with its environment
\citep[e.g.][]{blanton03a}, it is not suprising that galaxy color
would be more closely correlated with galaxy environment. [OII]
equivalent width more closely traces the instantaneous star-formation
rate of a galaxy, which can be dictated by short-term processes such
as minor mergers that are not inherently tied to the local density of
bright galaxies on somewhat larger scales, which we measure in the
DEEP2 data. Furthermore, $W_{\rm [OII]}$ is influenced by additional
factors such AGN activity \citep{yan05} which may further smear out
the correlation between environment and star-formation as traced by
$W_{\rm [OII]}$. Color, on the other hand, tracks the history of the
galaxy on longer time-scales and as such may be more strongly
correlated with the larger-scale environment of the galaxy as probed
by the DEEP2 survey.


\subsection{Dependence of Mean Environment on Luminosity}

Within the local universe there is striking evidence that the
correlation between environment and absolute magnitude is heavily
dependent on color. \citet{hogg03} find no luminosity dependence on
the mean environment of nearby blue galaxies. For the red population,
however, a strong luminosity dependence is observed, with the most
luminous red galaxies on average residing in increasingly dense
environments and with the intrinsically faintest local galaxies also
found to populate regions of greater density \citep{hogg03,
blanton03b}.

When attempting to examine the luminosity dependence of environment in
the DEEP2 survey, we must acknowledge the strong relationship between
mean color and luminosity in the data. At $M_B > -20.5$, the DEEP2
galaxy sample is dominated by blue galaxies including a population of
faint $(M_B > -20)$, very blue $((U-B)_0 < 0.4)$ objects (see Figure
\ref{cmd}). Therefore, as a fair attempt to disentangle the
environment dependencies on color and on luminosity at $z \sim 1$, we
compute the weighted mean galaxy overdensity as a function of
luminosity for red and blue galaxies separately. That is, the
individual overdensity values are still computed using the entire
galaxy sample, but the mean relations with luminosity are computed for
the red and blue samples independently. We divide the galaxy sample
into red and blue populations according to the color criterion defined
by \citet{willmer05}.

\begin{figure*}[h]
\begin{center}
\plotone{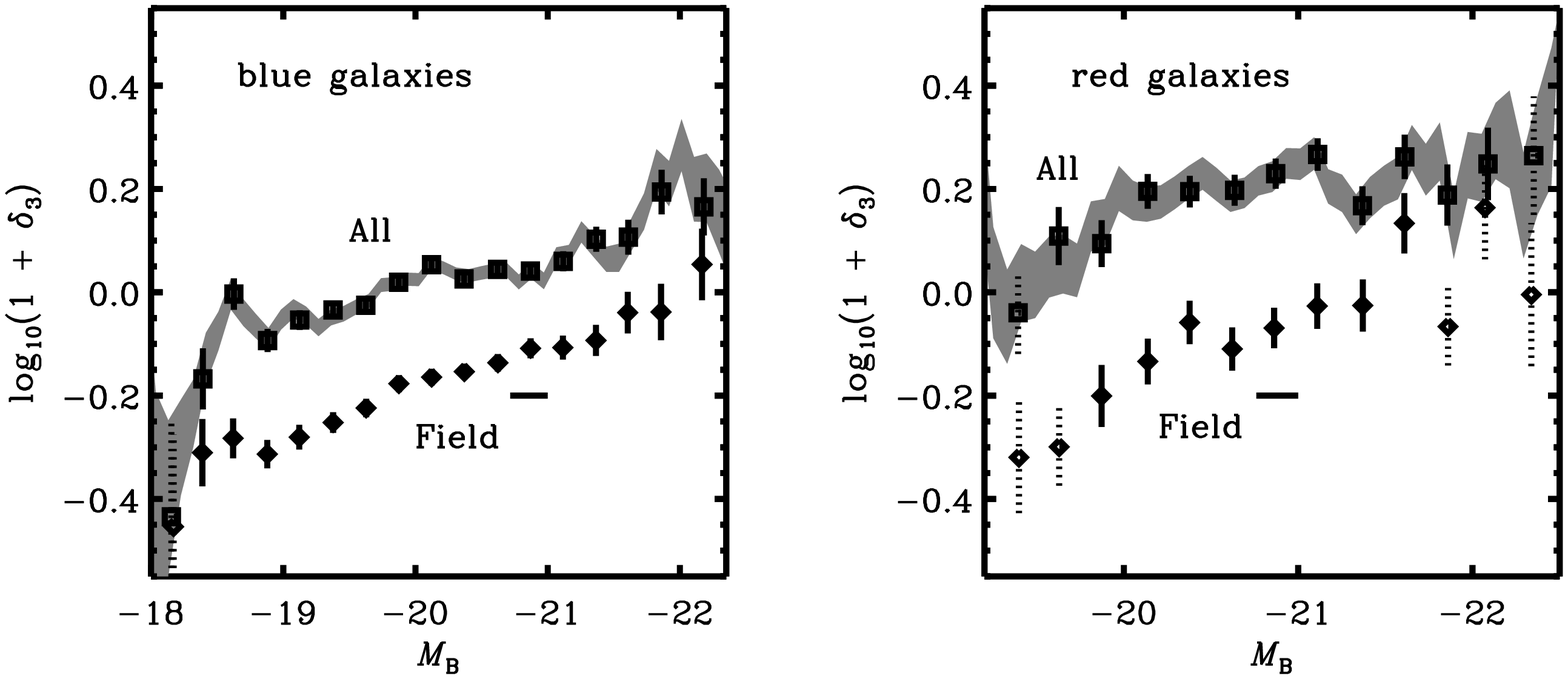}
\caption{(\emph{Left}) We plot the logarithm of the local galaxy
overdensity, $(1 + \delta_3)$, versus $B$-band absolute magnitude for
all blue galaxies in Sample C. The blue population is selected
according to the color division presented by \citet{willmer05}. We
measure the weighted mean dependence of environment on luminosity with
the mean computed using a sliding box of width given by the black dash
in the plot and weighted according to the inverse selection volumes,
$1/V_{\rm max}$. The grey region corresponds to the sliding 1-$\sigma$
uncertainties in this weighted mean. The square points give the
weighted mean environments and 1-$\sigma$ uncertainties in the means
computed in distinct bins of $M_B$, thereby avoiding the covariance
associated with the sliding mean. Means computed in bins containing
less than 50 galaxies have been drawn with dotted error bars to
signify that the uncertainty in these points has likely been
underestimated. The diamond points and error bars show the weighted
mean environments and 1-$\sigma$ uncertainties in the means computed
in indentical luminosity bins for the field galaxy population
solely. The field population is selected according to the galaxy group
finder of \citet{gerke05}, excluding all galaxies identified as group
members. (\emph{Right}) We make an identical plot, restricting the
sample to all red galaxies drawn from Sample C according to the color
division of \citet{willmer05}. The total number of blue galaxies is
12,120, while 2,094 comprise the red galaxy sample. Similar to studies
at $z \sim 0$, we find a strong trend between mean environment and
luminosity for red galaxies at $z \sim 1$. However, we also observe a
comparable trend with the blue galaxy population that is not found
locally.}
\label{environ_vs_amagBR}
\end{center}
\end{figure*}


Figure \ref{environ_vs_amagBR} shows the mean dependence of galaxy
overdensity on absolute $B$-band magnitude for the red and blue galaxy
populations in the data sample. For the red population, we find clear
evidence for luminosity dependence similar to that found locally; over
the entire luminosity range probed by the survey, the mean overdensity
for red galaxies increases with luminosity. Our results, however,
probe a significantly smaller portion of the galaxy luminosity
function than similar studies at low redshift; DEEP2 is unable to
study the faintest red galaxies at high redshift which do not make the
$R_{\rm AB} \le 24.1$ magnitude limit for the survey, while the most
luminous red galaxies observed in the SDSS are very rare, so few are
included in the much smaller DEEP2 survey volume.

Among the blue $z \sim 1$ population, we find a clear slope in the
relation between mean environment and $B$-band absolute magnitude,
with brighter blue galaxies found on average in regions of greater
overdensity. This trend strongly contrasts local studies by
\citet{hogg03} and \citet{blanton03b}, which find that the mean
environment of blue galaxies is independent of luminosity. If we fit
$\langle \log (1 + \delta_3)(M_B) \rangle$ as a linear function of
$M_B$, we find consistent slopes for the red and blue samples, with
values of $-0.85 \pm 0.35 \log(1 + \delta_3)/M_B$ and $-1.26 \pm 0.12
\log(1 + \delta_3)/M_B$, respectively (see Table \ref{fits_tab}).

Restricting our sample to the field population using the group finder
of \citet{gerke05}, we find that the dependence of mean environment on
$B$-band luminosity in the field shows a steeper slope $(-2.65 \pm
0.47 \log(1 + \delta_3)/M_B$ and $-1.93 \pm 0.15 \log(1 +
\delta_3)/M_B)$ for both the blue and red galaxy populations,
respectively. Red field galaxies, however, are rare at $z \sim 1$ and
thus the measured luminosity-environment trend among the red
population is rather noisy. Among the blue galaxy field sample,
though, the sample size is statistically significant. Within the blue
field population, the mean environment is at or below the mean survey
density $(\log{1 + \delta_3} = 0)$ over the entire luminosity range
probed.

\subsection{Predicting Environment from Galaxy Properties}

Results from \S5.2 indicate that galaxy color is more closely related
to environment than [OII] equivalent width is at $z \sim 1$. To study
this quantitatively, we compute the $1/V_{\rm max}$-weighted variance
$(\sigma^2)$ of the overdensity $(1 + \delta_3)$ measures and compare to
the variance computed relative to the mean overdensity as a function
of galaxy property $X$, $\sigma^2_{X}$, where

\begin{equation}
\sigma^2_{X} = \left\langle \left[ \log (1 + \delta_{3,i})  - 
\overline{\log (1 + \delta_3)} \left(X_i\right)  \right]^2 
\right\rangle .
\label{var_eqn} 
\end{equation}

This metric, which we use following \citet{blanton03b}, gives the
scatter about the mean relation between property $X$ and overdensity
and has the benefit of being independent of the differing units for
differing quantities $X$. In Table \ref{var_tab}, we present the
values of $(\sigma^2_X - \sigma^2)$ for each property examined in this
work and for each pair of properties.
The $1/V_{\rm max}$-weighted variance in the measured overdensity
values, $\sigma^2$, is 0.410 for the sample, and from the errors in
the overdensity values alone, we expect a scatter of 0.249. While the
local study by \citet{blanton03b} considered a slightly different set
of galaxy properties (rest-frame color, luminosity, and morphology),
our results are in agreement with their findings. For the DEEP2
sample, rest-frame $(U-B)_0$ color is the galaxy property most
predictive of environment, among those tested. As a pair, galaxy color
and luminosity show the greatest correlation with the observed galaxy
densities, explaining only $\sim 10\%$ of the variation.

\subsection{Environment as a Function of Color and
Luminosity}

As shown in \S5.3, the dependence of mean environment on luminosity is
very similar for both red and blue galaxy populations at $z \sim
1$. This suggests that the mean dependence of overdensity on
rest-frame color and absolute magnitude may be separated into two
functions:
\begin{equation}
\left\langle \log (1 + \delta_3) \left[ M_B, (U-B)_0 \right] \right\rangle 
= f\left[ \left(U-B\right)_0 \right] + g \left[ M_B \right]
\label{sep_eqn}
\end{equation}
where the mean dependence on galaxy color, $f[(U-B)_0]$, is given by
the $6^{\rm th}$-degree-polynomial fit to the observed correlation
with $(U-B)_0$ color as shown in Figure \ref{environ_vs_color} and
Table \ref{fits_tab}, and the mean dependence on luminosity is a
linear fit derived from Figure \ref{environ_vs_amagBR} in \S5.3 and
listed in Table \ref{fits_tab}. In this analysis, we limit the sample
to the regime $0.75 < z < 1.05$ where evolution in the measured
density values is small and the red galaxy population is least
affected by the survey target-selection.

\begin{figure*}[h]
\begin{center}
\plotone{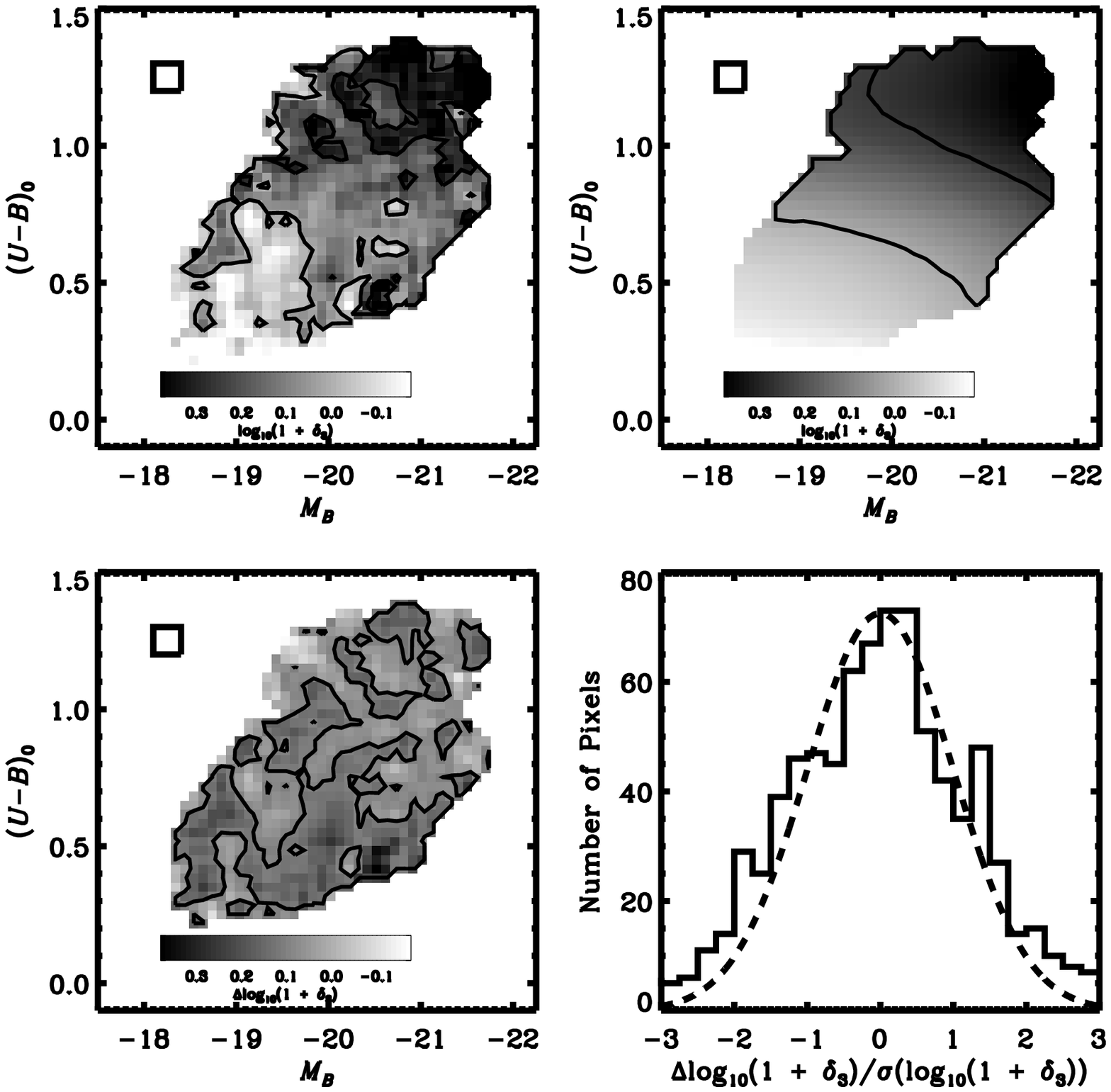}
\caption{(\emph{Top Left}) We display the $1/V_{\rm max}$-weighted
mean galaxy overdensity, $\log{(1 + \delta_3)}$, as a function of
galaxy color, $(U-B)_0$, and absolute magnitude, $M_B$, computed in a
sliding box of width $\Delta M_B = 0.3$ and height $\Delta (U-B)_0 =
0.1$. The size and shape of the box are illustrated in the upper
right-hand corner of the plot. Darker areas in the image correspond to
regions of higher average overdensity in color-magnitude space with
the scale given in the color-bar. The contours correspond to
overdensity levels of $\log{(1 + \delta_3)} = 0.0, 0.2, 0.4$. At
regions where the sliding box includes less than 20 galaxies, the mean
environment is not displayed. (\emph{Top Right}) The mean environment
as a function of color and magnitude as predicted by our separable
model, $\left\langle \log (1 + \delta_3 \left(M_B, (U-B)_0 \right) )
\right\rangle = \left\langle \log (1 + \delta_3 \left(M_B \right) )
\right\rangle + \left\langle \log (1 + \delta_3 \left((U-B)_0 \right) )
\right\rangle$ using the same sliding box. The overlayed contours
correspond to the same mean overdensity levels as traced in the
\emph{Top Left} image. (\emph{Bottom Left}) We display the residual
mean enviroment $\log{(1 + \delta_{3,{\rm data}})} - \log{(1 +
\delta_{3,{\rm model}})}$ as a function of color and magnitude
computed in the same sliding box. The contour plotted corresponds to a
residual of zero $\sigma$.
(\emph{Bottom Right}) We plot the distribution of residuals in units
of $\sigma(<\log \delta_3(M_B, (U-B)_0)>)$ for all regions where the
sliding box contains 20 or more galaxies. The distribution is roughly
Gaussian with $\sim 50\%$ of the positions having a residual with
absolute value greater than 1-$\sigma$. Overplotted for comparison is
a Gaussian distribution of amplitude 65 pixels and dispersion of
unity. The dependence of mean environment on rest-frame color and
luminosity at $z \sim 1$ is well represented by a separable function
of the form $\left\langle \log \delta_3 \left(M_B, (U-B)_0 \right)
\right\rangle = \left\langle \log \delta_3 \left(M_B\right)
\right\rangle + \left\langle \log \delta_3 \left((U-B)_0 \right)
\right\rangle$.}
\label{environ_model}
\end{center}
\end{figure*}

In the top-left panel of Figure \ref{environ_model}, we show the mean
environment as a function of both color and luminosity for galaxies
within $0.75 < z < 1.05$. The strong trend with galaxy color is
evident as well as the dependence on luminosity. Using the model given
in equation \ref{sep_eqn}, we predict the mean dependence of
environment on $(U-B)_0$ and $M_B$, as shown in the top-right panel of
Figure \ref{environ_model}. Within the errors of our measurements, the
model accurately predicts the mean overdensity over the entire
color-magnitude space. The bottom panels of Figure \ref{environ_model}
show the residuals between the data and the model, which are only
slightly non-Gaussian.

\section{Discussion}

\subsection{The Downsizing of Quenching}

Recent studies of the galaxy luminosity function at $z \sim 1$
\citep[e.g.][]{bell04,faber05} indicate that the build-up of stellar
mass at the high-mass end of the red sequence from $0 \lesssim z
\lesssim 1$ did not occur due to star formation within red
galaxies. Instead, these studies conclude that massive red galaxies
observed locally migrated to the bright end of the red sequence by a
combination of two processes: (1) the suppression of star formation in
blue galaxies (that is, the ``quenching'' of blue galaxies) and (2)
the merging of less-luminous, previously-quenched red galaxies. With
respect to the first scenario, \citet{faber05} discuss the growing
body of evidence suggesting that the typical mass at which a blue,
star-forming galaxy is quenched has decreased with time, the so-called
``downsizing of quenching''. The theory of downsizing has been
explored in the literature for quite some time \citep[see e.g.][and
references therein]{faber05}, with \citet{cowie96} first presenting
the concept in terms of a decline with redshift in the typical mass of
star-forming galaxies. While the process of quenching and the activity
of star formation are clearly closely related, the typical time-scale
for each may have a different dependence on mass. Here, we will focus
our discussion on the aspect of quenching within the larger picture of
downsizing.


The correlations between environment and galaxy properties at $z \sim
1$, as presented in this paper, add further observational support to
the proposed downsizing of quenching with redshift. In contrast to the
observed $z \sim 0$ trends, we find a clear dependence of mean
environment on luminosity for blue galaxies, with the brightest blue
galaxies residing in regions of greater overdensity. These bright blue
galaxies are also known to populate the high-mass end of the
color-magnitude blue cloud, with stellar masses of $\sim {\rm a\ few}
\times 10^{10}\ {\rm M}_{\sun}$ \citep{bundy06, cooper06c} and are
preferentially found in regions of high overdensity where quenching is
most likely to occur, the locations in models where it is driven by
galaxy mergers, the stripping of gas by hydrodynamical and tidal
effects, or AGN. Furthermore, those mechanisms which can cause the
total stellar mass of a large galaxy to decrease (e.g.\ galaxy
harassment) are only effective in the most massive clusters, which are
not represented within the DEEP2 sample (see \S 6.2). Therefore, it is
extremely unlikely that these galaxies can decrease in stellar mass
from $z \sim 1$ to $z \sim 0$. Thus, we conclude that the bright blue
galaxies in overdense environments at high redshift evolve (i.e., have
their star formation quenched) into members of the red sequence by $z
\sim 0$. As such, our results within the larger set of observations
suggest that the \emph{typical} entry point onto the red sequence via
quenching is fainter (i.e., less massive) at $z \sim 0$ than at $z
\sim 1$. That is, the luminosity dependence of the mean environment
for blue galaxies within the DEEP2 sample, when compared with the
local SDSS and recent DEEP2 results, is consistent with a ``downsizing
of quenching'' as discussed by \citet{faber05}, in the sense that
there exists a population of massive galaxies that was likely
undergoing quenching at $z \sim 1$, which has no high-mass counterpart
today. As part of an upcoming paper \citep{cooper06c}, we intend to
examine these trends in detail by including stellar mass
determinations alongside environment, color, and luminosity measures.

\subsection{The Roles of Field, Group, and Cluster Environments at
  High Redshift}

Using the group finder of \cite{gerke05}, we can associate each galaxy
in our analysis with either a cluster, a group, or the field. This
provides an interesting way to examine the contribution from bound
structures of different size to the environmental trends presented in
\S 5. The DEEP2 galaxy catalog is dominated by group and field
galaxies, with large clusters only scarcely represented. As
illustrated in Fig.\ \ref{environ_vs_color}, the relationship between
median environment and rest-frame color mirrors the dependence
observed for the mean, with the median relation negligibly affected by
cluster members. Within the survey volume (sample C), only $\sim 2\%$
of galaxies are found in clusters (i.e.\ groups with velocity
dispersions greater than $500\ {\rm km}/{\rm s}$ and $N_{\rm group} >
4$ observed members). For the red sequence alone, the contribution is
only slightly larger at $\lesssim 3\%$. On the other hand, field galaxies
represent $\sim 67\%$ of the total, with the remaining $\sim 30\%$
locked into galaxy groups. Such groups contain on average $\sim 3$
members in the survey sample, thus $\sim 5\ L_{*}$ members overall
given the $\sim 60\%$ sampling rate of the DEEP2 survey.

Knowing that cluster-sized objects within the DEEP2 survey volume make
a negligible contribution to our results allows us to place
constraints on the physical processes which may or may not have had a
role in establishing the color-environment and luminosity-environment
relationships. For example, in highly overdense environments numerous
mechanisms have been identified that can potentially alter the colors
of the cluster population, including ram-pressure stripping
\citep{gunn72, abadi99}, galaxy harassment \citep{moore96, moore98},
and global tidal interactions within the cluster potential
\citep{byrd90,valluri93}. Although such processes all modify the
star-formation histories of the galaxies on which they operate, such
environments are not present in the DEEP2 sample in any statistically
significant sense and the aforementioned processes do not operate
efficiently in lower-density environments such as groups. Thus, these
cluster-specific mechanisms alone cannot explain the established
environmental relationships observed at $z \sim 1$.

The interesting flip-side to the above discussion is that both
group-sized systems and galaxies in the field are abundant in the
DEEP2 galaxy catalog. Such sub-populations offer an alternative place
to look to understand the observed color- and luminosity-environment
trends. To disentangle the contribution from each, we first remove all
group and cluster members from the sample and then recompute each of
our main results. In Figure \ref{environ_vs_color}, we overplot the $z
\sim 1$ relation between mean overdensity and galaxy rest-frame
$(U-B)_0$ color for field galaxies alone (diamond symbols). We find
that for the field population only a weak trend of overdensity with
galaxy color persists, where red field galaxies are found to be only
marginally overdense relative to their blue counterparts. Thus, not
surprisingly, group-sized systems play a dominant role in establishing
the density contrast observed at the color bimodality within the
color-environment relation at $z \sim 1$.

While the difference in mean overdensity between red and blue galaxies
is reduced by excluding group members, the trend towards lower mean
overdensity at very blue colors is barely altered. The preservation of
this trend among the field population is not surprising given the
small fraction $(\sim 20\%)$ of galaxies at the bluest colors
$((U-B)_0 < 0.3)$ that are in groups. 
Based on their [OII] equivalent widths, these very blue galaxies have
the highest specific star-formation rates within the DEEP2 sample,
which indicates that their stellar populations are predominantly
composed of young stars. The nature of this interesting subpopulation
will be investigated more throughly by \citet{croton06}.

Having isolated the DEEP2 field galaxy population, we examine the
dependence of overdensity on rest-frame $B$-band absolute magnitude
for blue-cloud and red-sequence galaxies separately, as shown in
Figure \ref{environ_vs_amagBR} (diamond symbols, top and bottom
panels, respectively). Red-sequence field galaxies show a decrease in
density contrast relative to the full red sequence across the entire
magnitude range probed. Unfortunately, as such galaxies are rather
rare within the DEEP2 sample, the statistics we obtain are somewhat
noisy. Blue field galaxies, on the other hand, are greater in number
within DEEP2 and exhibit an interesting relationship between
environment and luminosity; when the contribution from group galaxies
is removed, their contrast in density also decreases. However, the
slope of this trend from faint to bright appears to have steepened
somewhat (see Fig.\ \ref{environ_vs_amagBR} and Table \ref{fits_tab}),
with the faintest blue field galaxies characteristically populating
survey regions with overdensities approaching those of voids and the
brightest blue field galaxies residing on average in environments of
mean survey density $(\log{\delta_3} = 0)$. 

This behavior is consistent with current observations of the local
galaxy population, where the characteristic luminosity of the blue
field population also scales with environment, at least in under- to
mean-density regions of the universe \citep[see Figure 7
of][]{croton05a}.
Thus in comparison to the results of \citet{croton05a}, at least
qualitatively, the blue sequence field population shows a similar
relationship with environment at both $z \sim 1$ and the present
day. This contrasts the bright blue group members, which through the
quenching of their star formation must transition onto the red
sequence where they are observed today in the local universe. This
leads us to propose that at $z \gtrsim 1$ quenching is a process that
dominates galaxy evolution down to systems of group size and not
below. Note that, although some red galaxies are found to reside in
low density environments at $z \sim 1$ they constitute only a small
portion of the field population \citep[$\lesssim 15\%$][]{gerke06}.
Conceivably, some of these galaxies could belong to groups missed by
our group finder or to fossil groups of high virial mass but with
underluminous satellite members. A more thorough study of the
relationship between environment and luminosity in group and field
galaxies in the SDSS or the 2dFGRS is truly needed to measure the
luminosity-environment trend among the blue field population locally,
thereby completing this picture.


\subsection{The Physics of Quenching}

One may naturally ask exactly what physical processes could have
occurred within both cluster \emph{and} group environments to explain
the observed color and magnitude relations with respect to overdensity
presented in this paper. We have already pointed out that many
mechanisms that dominate cluster-sized systems \emph{alone} are
unlikely to explain our results, as they do not operate efficiently in
the lower-mass halos where our results show they need to. We will now
discuss several processes that may be operating in this lower-mass
regime.

As shown by \citet{birnboim03}, \citet{keres05}, and
\citet{croton05b}, models of gas infall onto dark matter halos produce
a natural bimodal distribution in gas temperature, dividing cold
infalling gas from that which remains at the virial temperature to
form a static hot halo. The dominant component of gas in a system,
cold or hot, is found to change at a reasonably well defined dark
matter halo mass of approximately $2-3 \times 10^{11}\ {\rm M}_{\sun}$
out to at least $z \sim 6$. Interestingly, this mass is low enough to
include systems which are identified observationally as groups; within
the DEEP2 survey, galaxy groups are found to occupy dark matter halos
with masses both above and below this critical mass
\citep{coil06}. The above authors speculate that as a growing dark
matter halo passes this critical mass, the infalling cold gas supply
to the central galactic disk is shut off, and the galaxy will quickly
burn its remaining fuel and redden \citep[see also][]{dekel05}. Within
this picture, blue galaxies in the DEEP2 $z \sim 1$ sample are only
just approaching this critical halo mass and will still be
cold-accreting infalling gas from which star formation is fueled, even
for the brightest blue galaxies in the densest environments (Figure
\ref{environ_vs_amagBR}). However, dark matter halos continue to grow
at redshifts of $z < 1$, and thus these systems may eventually reach
the transition mass, at which point the infalling gas instead heats to
the virial temperature to build a quasistatic hot halo around the
galaxy. If there is no subsequent cold gas supply to the disk, then
star formation in such galaxies will ultimately become quenched, as
observed in the local universe.

The primary problem with this picture is that in systems more massive
than the critical halo mass, quasistatic x-ray emitting hot
atmospheres typically have central cooling times much shorter than the
age of the universe \citep[e.g.][]{fabian03}. The hot, dense central
regions must then cool, and the gas condensing out of the flow will
ultimately reach the galactic disk to feed the central galaxy
star-formation reservoir, in a similar way as the infalling
cold-accreting gas would when the system was much less massive. This
is the classical ``cooling flow problem'' and has been extensively
investigated in the literature over the past 30 or so years \citep[see
e.g.][]{fabian94, mcnamara02}. If left unabated, such unconstrained
star formation can keep even the most massive galaxy in the ``blue
cloud'' \citep{croton05b}, in stark contrast to what is
observed. Because of this, the transition from cold to hot accretion
for a galaxy halo is unlikely to be able to account for the observed
quenching of $z \sim 1$ bright blue galaxies alone. Many authors have
thus proposed AGN as an energetically feasible mechanism with which
cooling gas can be suppressed \citep[e.g.][and references
therin]{binney01, zanni05, sanderson05}. \cite{croton05b} demonstrate
that if one assumes a low energy AGN heating source fed from the
quiescent hot halo surrounding the galaxy (which, from above, can form
when the host dark matter halo reaches a mass of $\sim 10^{11} {\rm
M}_{\sun}$), then the cooling flow gas can be effectively stopped,
thereby quenching star formation and pushing the galaxy from the blue
cloud and onto the red sequence.

Other group-scale processes may also contribute to the quenching of
star formation. In addition to the mechanisms just discussed, galaxy
mergers, which preferentially occur in galaxy groups
\citep{cavaliere92, navarro87}, may be capable of triggering events
which remove cold gas from the galactic disk to shut off star
formation, at least for some (perhaps extended) period of time. Such
events can then potentially convert members of the blue cloud into
red-sequence galaxies. Early simulation work showed that mergers can
disrupt galactic disks yielding merger remnants with
surface-brightness profiles and density distributions similar to those
of elliptical galaxies \citep{toomre72, barnes89, barnes92}. Along
with a morphological transformation, merger events in systems rich in
cold gas can also produce starbursts from which strong galactic
super-winds flow \citep{cox05, benson03}, and merger-induced cold gas
accretion on to central super-massive black holes can trigger
energetic quasar winds which are capable of sweeping all gas from
their hosts \citep{murray05, hopkins05, robertson05, dimatteo05}. In
this way, a merger may initiate additional processes which result in a
depletion of the star forming gas reservoir, leading ultimately to a
quenching of the galaxies star formation if no addition cold gas
supply is present.


We conclude this section by mentioning an additional process that may
be relevant to the present discussion, but may not be described as an
environment-dependent effect. Secular evolution (or secular
bulge-building) is another means of triggering morphological evolution
in a galaxy \citep[e.g.][and references therein]{zhang96, kormendy04,
kormendy05}. Following secular evolution, subcomponents of a galaxy
interact internally through the transfer of angular momentum along a
galactic bar (or spiral arms), from which matter can funnel inwards
towards the galactic center. For example, \citet{noguchi99} suggests
that the bulges of disk galaxies may result from the accumulation of
massive clumps that formed within the galactic disk and migrated
inward via dynamical friction. Such instabilities provide a channel
through which cold gas in the disk can transfer away its angular
momentum to reach the central regions of the galaxy, potentially
providing a source of fuel to the central black hole to trigger an AGN
outflow, as described above. The results presented within this paper,
however, suggest that the quenching of blue galaxies is an
evolutionary process that occurs preferentially in overdense
environments. Thus, it appears as if secular evolution cannot be the
\emph{only} process responsible for the trends with environment
observed at $z \sim 1$.


\subsection{Evolution of Environment Trends from \boldmath$z \sim 1$ 
to \boldmath$z \sim 0$}

While a clear relationship between rest-frame color and environment is
already in place by a redshift of unity, and comparison to local
samples indicates that star formation is being suppressed in more
massive galaxies at $z \sim 1$ than at $z \sim 0$, we are
unfortunately not yet able to precisely quantify the differences
between the two populations or the strength of this evolution. Such a
quantitative analysis requires establishing the local and $z \sim 1$
galaxy samples on equal footing, for example, by measuring galaxy
colors and luminosities in identical passbands and measuring
environments using identical techniques and with equivalent galaxy
samples. While beyond the scope of this current paper, future work
utilizing the DEEP2 and SDSS data sets \citep{cooper06d} will address
this issue.

It is possible, however, to combine the environment measurements at $z
\sim 1$ and $z \sim 0$ into a general picture of how the overall
galaxy population evolves. Here, we discuss the results presented in
this work and similar low-$z$ studies \citep[e.g.,][]{hogg03,
blanton03b}, in light of both the larger framework of clustering
evolution and the need for quenching mechanisms that operate most
efficiently in group- and cluster-sized systems. The color-environment
relations at low- and high-redshift epochs show amazing similarity in
general. While quantifying the evolution in this relationship will be
addressed in another paper, we speculate that the contrast in mean
overdensity between red and blue galaxies increases from $z \sim 1$ to
$z \sim 0$ as blue galaxies in overdense environments are
preferentially quenched and join the red sequence and as
already-quenched red galaxies continue to cluster into increasingly
overdense environments.

This larger picture is also consistent with the observed trends
between environment and luminosity for both red and blue galaxy
populations. 
Unlike similar $z \sim 0$ studies, the DEEP2 survey unfortunately does
not sample the faint end of the red galaxy luminosity function due to
the $R_{AB}$ magnitude limit of the survey. At bright magnitudes,
however, we find a trend towards increasing mean environment with
luminosity for red galaxies similar to that found locally. The
increasingly steep slope to the $z \sim 0$ luminosity-environment
relation at the very bright end of the red sequence \citep{hogg03,
blanton03b} is not seen in our higher-redshift results. To build such
bright red massive galaxies requires cluster environments, which are
relatively rare at $z \sim 1$ but increase in number in a
highly-evolved clustering distribution as seen locally.

For blue galaxies in all environments, the $z \sim 1$ relationship
between mean overdensity and luminosity is significantly different
from that seen at $z \sim 0$, with a strong slope such that bright
blue galaxies are preferentially found in regions of higher
overdensity and faint blue galaxies in underdense regions. In
contrast, \citet{croton05b}, suggest that the blue \emph{field} galaxy
population at low redshift shows a luminosity-environment trend quite
similar to the $z \sim 1$ relation presented in this paper. Evolution
such as this could be expected, if as the clustering of the galaxy
distribution evolves, the bright blue galaxies in overdense regions at
$z \sim 1$ are preferentially quenched, while the faint blue cluster
population becomes increasingly numerous from satellites first falling
into denser regions. This evolution at the bright and faint end of the
blue population would cause the trend between environment and
luminosity along the blue cloud to flatten, thereby explaining the $z
\sim 0$ results of \citet{hogg03} and \citet{blanton03b}.

In future work, we will attempt to clarify the exact physical
processes responsible for the quenching of blue, star-forming galaxies
by studying the relationship between environment, AGN activity, and
galaxy morphology within the DEEP2 data set \citep{cooper06c}. There
the multiwavelength data of the Extended Groth Strip (EGS) will be
invaluable in identifying the AGN population, and high-resolution
HST/ACS imaging will be critical for classifying galaxy morphologies.

\section{Conclusions}

In this paper, we present the first study of the relationships between
galaxy properties and environment at $z \sim 1$ to cover a broad and
continuous range of environment from groups to weakly-clustered field
galaxies. Using a sample of galaxies drawn from the DEEP2 Galaxy
Redshift Survey, we estimated the local overdensity about each galaxy
according to the projected $3^{\rm rd}$-nearest-neighbor surface
density. We study the relationships between environment and galaxy
color, luminosity, and [OII] equivalent width. Our principal results
are:

\begin{enumerate}

\item We find a strong dependence of mean environment on rest-frame
$(U-B)_0$ color (Fig.\ \ref{environ_vs_color}), with blue galaxies on
average occupying regions of lower density than red galaxies. The
observed trend with galaxy color echoes the results of studies of
nearby galaxies \citep[e.g.][]{hogg03, blanton04}; all features of
the global correlation between galaxy color and environment measured
at $z \sim 0$ are found to be already in place at $z \sim 1$.

\item For the red galaxy population at $z \sim 1$, we see clear
evidence for a dependence of mean overdensity on luminosity, mirroring
results from local studies (Fig.\ \ref{environ_vs_amagBR}). Unlike the
SDSS, however, DEEP2 does not probe the faint red galaxy population
and includes relatively few luminous red galaxies due to the smaller
survey volume, rendering the strongest effects seen locally
undetectable.

\item While low redshift studies find the mean environment of blue
galaxies to be independent of luminosity, we find a strong increase in
local density with luminosity for blue galaxies at $z \sim 1$ (Fig.\
\ref{environ_vs_amagBR}). Restricting to the blue field population in
the DEEP2 survey, the dependence of mean overdensity on $B$-band
luminosity persists, with the mean environment at or below the mean
survey density over the luminosity range probed.


\item The pair of galaxy properties which best predict environment at
$z \sim 1$ are galaxy color and luminosity (Table
\ref{var_tab}). Additionally, while we do measure a modest difference
in luminosity dependence for the mean environments of blue and red
galaxies, we show that the dependence of environment on $M_B$ and
$(U-B)_0$ is well represented by a separable function (Fig.\
\ref{environ_model}).

\item The mechanisms of ram-pressure stripping, galaxy harassment, and
global tidal interactions, which preferentially occur in clusters,
cannot alone explain the observed relationships between galaxy
properties and environment at $z \sim 1$.

\item Our results are consistent with a downsizing of the
characteristic mass or luminosity at which the quenching of a galaxy's
star formation becomes efficient. We discuss how quenching appears to
be a process that must operate efficiently in both cluster {\em and}
group environments for consistency between our results and those at
redshift $z \sim 0$. AGN heating of cooling gas may satisfy this
requirement, however the exact quenching mechanism still remains an
open question.

\end{enumerate}


\acknowledgments This work was supported in part by NSF grant
AST00-71048. J.A.N., D.S.M., and A.L.C.\ acknowledge support by NASA
through Hubble Fellowship grants HST-HF-01165.01-A, HST-HF-01163.01-A,
and HST-HF-01182.01-A, respectively, awarded by the Space Telescope
Science Institute, which is operated by AURA Inc.\ under NASA contract
NAS 5-26555. S.M.F.\ would like to acknowledge the support of a
Visiting Miller Professorship at UC-Berkeley. M.C.C.\ thanks Michael
Blanton, David Hogg, and Guinevere Kauffmann for helpful discussions
that improved this work.

We also wish to recognize and acknowledge the highly significant
cultural role and reverence that the summit of Mauna Kea has always
had within the indigenous Hawaiian community. It is a privilege to be
given the opportunity to conduct observations from this mountain.


\pagebreak

\begin{deluxetable}{l c c c c}
\tablewidth{0pt}
\tablecolumns{5}
\tablecaption{\label{var_tab} Galaxy Samples }
\tablehead{Sample & Redshift Range & Edge Distance & 
$W_{\rm [OII]}$ &\# of Galaxies }
\startdata
Sample A & $0 < z < 2$ & $\cdots$ & $\cdots$ & 23,004 \\
Sample B & $0.75 < z < 1.35$ & $\cdots$ & $\cdots$ & 18,977 \\
Sample C & $0.75 < z < 1.35$ & $> 1 h^{-1}\ {\rm Mpc}$ & $\cdots$ &
14,214 \\
Sample D & $0.75 < z < 1.35$ & $> 1 h^{-1}\ {\rm Mpc}$ &
$\sigma_{W_{\rm [OII]}} < 10$\AA & 11,250 \\
Sample E & $0.75 < z < 1.05$ & $> 1 h^{-1}\ {\rm Mpc}$ & $\cdots$ &
9,567 \\ 
\enddata

\tablecomments{We list all of the galaxy samples used in this
work. For each sample, we give the redshift range over which galaxies
are selected, any restrictions applied regarding the distance to the
nearest survey edge (see \S 4.2) or according to error in equivalent
width (see \S 3.2). In the final column, we detail the number of
galaxies included in each sample. All samples are restricted to
sources with accurate redshift determinations (Q=3 or Q=4 redshifts as
defined by \citet{faber06}).}

\label{sample_tab}
\end{deluxetable}


\begin{deluxetable}{c c c c c c c}
\tablewidth{0pt}
\tablecolumns{7}
\tablecaption{\label{var_tab} Polynomail Fits to Mean Relations}
\tablehead{ & $a_0$ & $a_1$ & $a_2$ & $a_3$ & $a_4$ & $a_5$ }
\startdata
$(U-B)_0$ & -0.658 & 4.865 & -15.126 & 21.662 & -13.762 & 3.175 \\
$M_B [{\rm blue}]$ & -1.249 & -0.063 & $\cdots$ & $\cdots$ & 
$\cdots$ & $\cdots$ \\
$M_B [{\rm red}]$ & -0.826 & -0.049 & $\cdots$ & $\cdots$ & 
$\cdots$ & $\cdots$ \\
\enddata

\tablecomments{We list the coefficients for the polynomial fit to the
mean environment versus color relation (see Fig.\
\ref{environ_vs_color}. Also listed are the coefficients for the
linear fits to the mean environment versus absolute magnitude
relations for red and blue galaxies (see Fig.\
\ref{environ_vs_amagBR}). For each, the functional form of the fit is
given by $f(x) = a_0 + a_1 x + a_2 x^2 + \cdots$.}

\label{fits_tab}
\end{deluxetable}

\begin{deluxetable}{c c c c c}
\tablewidth{0pt}
\tablecolumns{5}
\tablecaption{\label{var_tab} Correlation of Properties and
Environment} 
\tablehead{Galaxy Property & $(\sigma^2_X - \sigma^2)$ & 
$(U-B)_0$ & $W_{\rm [OII]}$ & $M_B$ }
\startdata
$(U-B)_0$ & -0.0076 & $\cdots$ & -0.0075 & -0.0160 \\
$W_{\rm [OII]}$ & -0.0019 & -0.0075 & $\cdots$ & -0.0147  \\
$M_B$ & -0.0031 & -0.0160 & -0.0147 & $\cdots$ \\
\enddata
\tablecomments{For each galaxy property studied in this paper, the
second column gives the values of $\sigma^2_X - \sigma^2$. The entries
in the right three columns give the difference in variance,
$\sigma^2_{X,Y} - \sigma^2$, where $\sigma^2_{X,Y}$ is the variance
about the mean relations for the pair of properties $X$ and $Y$. In
this table, lower values indicate that the galaxy property or pair of
properties are a better predictor of environment. Rest-frame color,
$(U-B)_0$, is the single best predictor of environment among the
properties studied, while the combination of color and $B$-band
luminosity are the most strongly correlated pair of properties with
environment, on average.}
\label{var_tab}
\end{deluxetable}

\end{document}